\documentclass[a4paper,11pt]{article} 
\usepackage{jheppub}

\usepackage{graphicx}

\usepackage{epsfig}

\usepackage{color} 
\usepackage{amssymb}
\usepackage{amsmath}
\usepackage{doi}
\usepackage{dsfont}
\usepackage{subfloat}
\usepackage{subfig}
\usepackage{float}
\usepackage{mathtools}
\usepackage{verbatim}
\usepackage{booktabs}
\usepackage{slashed}
\usepackage[normalem]{ulem} 
\usepackage{xspace}

\usepackage[dvipsnames]{xcolor}

\newcommand{\mr}{\mathrm}
\newcommand{\mc}{\mathcal}

\newcommand{\hjjj}{H+3~\text{jets}}
\newcommand{\hhjj}{HH+2~\text{jets}}
\newcommand{\vbfhh}{VBF~$HH$}

\newcommand{\vvjj}{VV+2~\text{jets}}

\newcommand{\muf}{\mu_\mr{F}}
\newcommand{\mur}{\mu_\mr{R}}
\newcommand{\xif}{\xi_\mr{F}}
\newcommand{\xir}{\xi_\mr{R}}

\newcommand{\VBFNLO}{{\tt{VBFNLO}}}

\newcommand{\MGAMCNLO}{{\tt{MadGraph5\_aMC@NLO}}} 
\newcommand{\MADGRAPH}{{\tt{MadGraph}}}

\newcommand{\PBOX}{{\tt{POWHEG~BOX}}}
\newcommand{\POWHEGBOXVV}{{\tt{POWHEG~BOX~V2}}}

\newcommand{\PYTHIA}{{\tt{PYTHIA}}}

\newcommand{\PYTHIAE}{{\tt{PYTHIA8}}}
\newcommand{\HERWIG}{{\tt{HERWIG}}}
\newcommand{\HERWIGS}{{\tt{HERWIG7}}}
\newcommand{\VINCIA}{{\tt{Vincia}}}

\newcommand{\provbfhh}{\texttt{proVBFHH}\xspace}
\newcommand{\provbfh}{\texttt{proVBFH}\xspace}

\newcommand{\beq}{\begin{equation}}
\newcommand{\eeq}{\end{equation}}

\newcommand{\bea}{\begin{eqnarray}}
\newcommand{\eea}{\end{eqnarray}}

\newcommand{\yjstar}{y_{j_3}^{\star}}
\newcommand{\mjjtag}{m_{jj}^{\mr{tag}}}
\newcommand{\yjjtag}{y_{jj}^{\mr{tag}}}

\newcommand{\ghhvv}{g_{HHVV}}
\newcommand{\ghvv}{g_{HVV}}
\newcommand{\ghhh}{g_{HHH}}
\newcommand{\lamhhh}{\ghhh}
\newcommand{\klam}{\kappa_\lambda}
\newcommand{\khhvv}{\kappa_{2V}}
\newcommand{\khvv}{\kappa_{V}}

\newcommand{\vfin}{\mc{V}_\mr{fin}}

\newcommand{\ed}{\end{document}}

\title{Precision tools for the simulation of double-Higgs production via vector-boson fusion}
\preprint{CERN-TH-2025-030}

\newcommand{\TUBaff}{Institute for Theoretical Physics, University of T\"ubingen, Auf der Morgenstelle 14,
72076 T\"ubingen, Germany}
\newcommand{\CERNaff}{CERN, Theoretical Physics Department, CH-1211 Geneva 23, Switzerland}

\author[a]{Barbara J\"ager,}%
\author[b]{Alexander Karlberg,}%
\author[a]{Simon Reinhardt}%

\affiliation[a]{\TUBaff}
\affiliation[b]{\CERNaff}

\date{Received: date / Accepted: \today}

\abstract{ We present two precision tools for the simulation of Higgs-pair
  production via vector-boson fusion in the kappa framework for the
  parameterization of non-standard Higgs couplings.
A new implementation of the process is developed in the framework of
the \PBOX{} program that can be used to provide predictions at the
next-to-leading order (NLO) of QCD matched to parton showers (PS).  In
addition, the existing \provbfhh program for the computation of
next-to-next-to-leading order (NNLO) QCD and
next-to-next-to-next-to-leading order QCD corrections is extended to
account for values of the Higgs couplings different from the
expectation of the Standard Model. 
We systematically compare and analyse predictions obtained with the
two programs and find that the NLO+PS predictions provide a good
approximation of the NNLO results for observables of the tagging jets
and Higgs bosons. The results turn out to be very sensitive to the
values of the modified Higgs couplings. Finally we study the
non-factorizable NNLO QCD corrections to the process in the presence
of anomalous couplings. We find that the size of the non-factorizable
corrections is very sensitive to the anomalous couplings.

  }

\keywords{QCD, Parton Shower, NLO, Matching, Higgs
  \\[4em]
  \textit{For the purpose of Open Access, the authors have applied a CC BY
  public copyright licence to any Author Accepted Manuscript (AAM)
  version arising from this submission.}
}

\begin{document}

\maketitle

\section{Introduction}
After the discovery of the Higgs boson by the
ATLAS~\cite{ATLAS:2012yve} and CMS~\cite{CMS:2012qbp} experiments at
the CERN Large Hadron Collider (LHC) in 2012 Higgs physics has entered
a precision era.
The production of a Higgs boson as predicted by the Standard Model (SM) has been measured in various production modes. 
No significant indications for physics beyond the SM (BSM) have been identified, and all measurements so far are compatible with the spin-zero, CP-even nature of the SM Higgs boson. To learn more about the nature of this particle,  a determination of the Higgs self couplings remains to be achieved, as these are intimately related to the shape of the Higgs potential. Such measurements are ideally performed in processes involving the pair production of two Higgs bosons. While the largest production rates are expected for the inclusive Higgs pair production process that predominantly proceeds via gluon fusion, the purely electroweak (EW) vector-boson fusion (VBF) channel, $pp\to\hhjj$,  exhibits smaller production rates yet better means for a selection of signal events via the characteristic tagging jets that accompany the Higgs bosons in the final state. 
A quantitative understanding of this channel is thus as important as
are flexible tools that can be used in experimental analyses and
phenomenological studies.

The ATLAS and CMS experiments have performed a series of searches for
Higgs-pair production both in inclusive setups and in the VBF channel
(see \cite{ATLAS:2024ish,CMS:2024ymd} for recent combinations of
experimental results), deriving bounds on the triple Higgs coupling $\ghhh$ as
well as the quartic Higgs-to-weak boson coupling $\ghhvv$.
All results are so far compatible with
the SM predictions.
Due to the low production cross section Higgs pair production has yet
to be discovered at the LHC, but it is expected to be so at the
upcoming HL-LHC.

The relevance of the \vbfhh{} process for a determination of the
triple Higgs coupling was first discussed in~\cite{Baglio:2012np}, and
its sensitivity to the quartic $HHVV$ coupling was explored
in~\cite{Bishara:2016kjn}. In the past there has also been interest in
studying the sensitivity of the \vbfhh{} process to specific
scenarios beyond the
SM~\cite{Boudjema:2001ii,Moretti:2004dg,Moretti:2004wa}, and the
discriminating power of VBF versus the gluon fusion background has
been studied in Refs.~\cite{Dolan:2013rja,Dolan:2015zja}. The
next-to-leading order (NLO) QCD corrections to the SM process are
available in the parton-level Monte-Carlo program
\VBFNLO{}~\cite{Arnold:2008rz,Baglio:2011juf,Baglio:2012np,Baglio:2024gyp},
and the NLO corrections have also been studied in the context of the
two-Higgs doublet model~\cite{Figy:2008zd}, 
 and in~\cite{Anisha:2024ryj} using Higgs Effective Field Theory for double and triple Higgs production via VBF.  NLO-QCD results matched
to a parton shower (PS) as obtained with the multi-purpose program
\MGAMCNLO{}~\cite{Alwall:2014hca} have first been presented
in~\cite{Frederix:2014hta}.
The leading factorizable next-to-next-to-leading order (NNLO) QCD
corrections for Higgs pair production via VBF were presented for
inclusive predictions in Ref.~\cite{Ling:2014sne}. The fully
differential predictions at this order were since computed using the
projection-to-Born method~\cite{Cacciari:2015jma,Dreyer:2018rfu} and
complemented by the inclusive next-to-next-to-next-to-leading order
(N$^3$LO) QCD calculation in~\cite{Dreyer:2018qbw}. While the NNLO-QCD
corrections were found to be significant in some regions of phase
space, yet higher orders of QCD were found to be very small. NLO-EW
corrections, on the other hand, can be quite pronounced in some
kinematic regions~\cite{Dreyer:2020xaj}.  Non-factorizable corrections
that are colour-suppressed but formally of the same order in the
strong coupling as the dominant factorizable NNLO-QCD corrections have
been found to be negligible after selection cuts typical for VBF
measurements~\cite{Dreyer:2020urf}.

The QCD corrections of
Refs.~\cite{Dreyer:2018rfu,Dreyer:2018qbw,Dreyer:2020urf} have been
implemented in the public \provbfhh program, available from
\href{https://github.com/alexanderkarlberg/proVBFH}{https://github.com/alexanderkarlberg/proVBFH}
together with the \provbfh
program~\cite{Cacciari:2015jma,Dreyer:2016oyx} for the related VBF
single Higgs production process.

In this article we will explore the impact of QCD corrections and PS
effects on observables of immediate relevance for the extraction of
the triple Higgs and the $HHVV$ coupling from measurements of the
\vbfhh{} process. To this end we present a new and public
implementation of the \vbfhh{} process in the
\PBOX{}~\cite{Alioli:2010xd}, a tool for the matching of NLO-QCD
corrections with PS generators using the POWHEG
formalism~\cite{Nason:2004rx,Frixione:2007vw}. In order to facilitate
a comparison of predictions with existing experimental results we have
also implemented the so-called {\em kappa
  framework}~\cite{Hagiwara:1993ck,LHCHiggsCrossSectionWorkingGroup:2012nn}
which accounts for new physics effects in the Higgs couplings in a
generic way. These couplings have also been fully implemented in the
\provbfhh{} program, providing predictions with modified couplings up
to N$^3$LO in the QCD coupling. The anomalous couplings have been
implemented in both the factorizable and non-factorizable corrections
in \provbfhh{}. Besides providing very fast inclusive cross section
predictions the program was also used to cross-check our \PBOX{}
implementation.

To analyse the capabilities of our new \PBOX{} implementation we
present two studies. First, we provide predictions at NLO+PS accuracy
with a number of widely used parton showers, namely
\PYTHIAE{}~\cite{bierlich2022comprehensiveguidephysicsusage},
\VINCIA{}~\cite{Fischer:2016vfv} and \HERWIGS{}~\cite{Bewick:2023tfi},
and compare them to the fixed-order NNLO-QCD predictions. We stress
that although NLO+PS predictions for the \vbfhh{} process can in
principle be obtained with \MGAMCNLO{}, it can at the moment only be
reliably matched to \HERWIGS{} as has been observed for other VBF
processes~\cite{Ballestrero:2018anz,Jager:2020hkz}. Our work therefore
overcomes an important obstruction in obtaining NLO accurate
predictions matched to the widely used \PYTHIAE{}-family of
showers. In general we find that the NLO+PS predictions provide a
reasonable approximation of the NNLO-QCD predictions for inclusive
observables.

Secondly we compute predictions at NLO+PS with modified Higgs
couplings for a number of distributions. Using values of the couplings
consistent with current experimental bounds show huge distortions
compared to the SM, highlighting the sensitivity of the \vbfhh{}
channel to these couplings.

Finally we study for the first time the interplay between anomalous
couplings and the higher order QCD corrections. In particular we find
that the non-factorizable corrections are extremely sensitive to the
anomalous corrections, and that enhancements of up to $40\%$ can be found
for even small coupling modifications. The factorizable corrections on
the other hand are almost completely independent of the anomalous
couplings, as expected.

The paper is structured as follows: We provide some details on the
respective calculations and tools in Sec.~\ref{sec:implementation} and
then present phenomenological results in Sec.~\ref{sec:pheno}. Finally
we conclude in Sec.~\ref{sec:conclusions}.

\section{Details of the implementation}
\label{sec:implementation}
EW Higgs pair production in association with two jets involves two
types of contributions: First, VBF topologies that are dominated by
the scattering of two (anti-)quarks via the $t$-channel exchange of
massive weak gauge bosons that subsequently emit two Higgs
bosons. Second, Higgs-strahlung contributions that are due to the
annihilation of a quark-anti-quark pair resulting in an $s$-channel
weak gauge boson that subsequently results in an on-shell Higgs pair
and a weak boson further decaying hadronically. While VBF- and
Higgs-strahlung topologies contribute to $\hhjj$ final states at the
same order $\alpha^4$ in the electroweak coupling, they give
rise to very different kinematic features of the jets. This can be
exploited to experimentally extract samples dominated by either
topology. In particular, VBF events are characterized by two so-called
{\em tagging jets} in the forward and backward regions of the detector
with large rapidity separation, $\yjjtag$, and invariant mass,
$\mjjtag$. When selection cuts typical for VBF analyses at the LHC are
imposed, the contribution of non-VBF topologies to the EW $\hhjj$
fiducial cross section was found to be at the sub-percent
level~\cite{Dreyer:2020xaj}. In this work we will thus focus on the
VBF production mode.
For our new \PBOX{} calculation we will furthermore assume that
contributions involving colour exchange between upper and lower
fermion lines in the VBF contributions are negligible. Within this
{\em VBF approximation} QCD corrections to upper and lower quark lines
decouple. The validity of this factorized approximation has been
verified at percent-level in Ref.~\cite{Dreyer:2020urf}. When
neglecting Higgs-strahlung contributions it is furthermore exact at
NLO-QCD.

\begin{figure}[t!]
\centering
\includegraphics[width=0.95\textwidth]{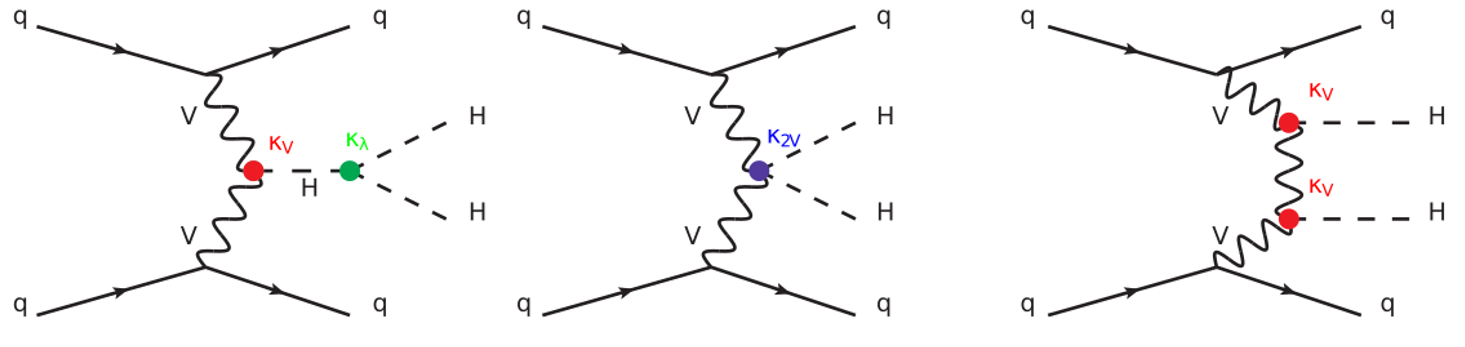}
\caption{\label{fig:diagrams} 
Representative Feynman diagrams contributing to \vbfhh{} production. 
}
\end{figure}

While the VBF-induced single Higgs production process has already been
extensively explored at the LHC, Higgs pair production is more
difficult to access due to the small associated cross
section. However, the \vbfhh{} process is of prime relevance for a
determination of Higgs couplings that cannot be accessed at tree level
in single-Higgs production processes, in particular the trilinear
Higgs coupling, $\lamhhh$, and the quartic coupling $\ghhvv$ between
Higgs and massive weak bosons $V=W^\pm, Z$.
A simple prescription to parameterize deviations from the SM, the
so-called {\em kappa framework}, has been suggested
in~\cite{Hagiwara:1993ck,LHCHiggsCrossSectionWorkingGroup:2012nn}. Within
this framework a coupling modifier $\kappa_i$ is defined as the ratio
of a coupling $c_i$ to the corresponding SM value $c_i^\mr{SM}$.  In
particular, we will express couplings of the Higgs boson entering the
VBF-induced $\hhjj$ process as \beq \lamhhh = \klam \cdot
\lamhhh^\mr{SM}\,, \quad \ghhvv = \khhvv \cdot \ghhvv^\mr{SM}\,, \quad
\ghvv = \khvv \cdot \ghvv^\mr{SM}.  \eeq Here, $\ghvv$ denotes the
trilinear coupling of a Higgs boson to two massive weak bosons. In
contrast to $\lamhhh$ and $\ghhvv$ this coupling is accessible in
single-Higgs production processes. However, it also enters the
VBF-induced Higgs-pair production process via diagrams where a single
Higgs boson couples to two weak bosons exchanged in the $t$-channel
and thus appears in our calculation. Representative LO Feynman
diagrams with anomalous couplings are shown in
Fig.~\ref{fig:diagrams}.

While the kappa framework does not constitute a fully unitarised
extension of the SM, it is often applied in experimental analyses
because of its simplicity. To allow experimentalists to nonetheless
take advantage of tools providing predictions of high accuracy, we
have extended the inclusive version of the \provbfhh program to
account for non-SM values of $\klam, \khhvv, \khvv$.  Additionally, we
have prepared an implementation of the \vbfhh{} process in the
context of the \POWHEGBOXVV{} (which we will just call the \PBOX{} in
the rest of the paper) that also offers the option for non-SM values
of the above-mentioned Higgs couplings. We stress that these two
implementations are completely independent, the \PBOX{} relying on
matrix elements and \provbfhh{} on structure functions, and that the
agreement between the codes therefore provides a very strong
cross-check.

We will discuss the extension of the \provbfhh program and the new
\PBOX{} implementation below.

\subsection{Extension of \provbfhh}
The \provbfhh
program~\cite{Dreyer:2018rfu,Dreyer:2018qbw,Dreyer:2020urf} is a tool
for the computation of NNLO- and N$^3$LO-QCD corrections to \vbfhh{}
production using the VBF approximation. From \texttt{v1.1.0} it can
also provide non-factorizable corrections in the eikonal
approximation. At NNLO it is fully differential whereas the
N$^3$LO-QCD corrections are inclusive in the jet kinematics, but fully
differential in the Higgs boson momenta. The code itself makes use of
some general features of the \PBOX{}, a modified implementation of the
VBF $\hjjj$ process in the \PBOX~\cite{Jager:2014vna} and of the
phase-space parameterization of the \vbfhh{} process as implemented in
\VBFNLO~\cite{Baglio:2012np}.  The anomalous couplings have been
implemented in \texttt{v2.1.0} and can be obtained from
\href{https://github.com/alexanderkarlberg/proVBFH}{https://github.com/alexanderkarlberg/proVBFH}.
 
For the purpose of studying the sensitivity of the \vbfhh{} process to
Higgs couplings while fully taking QCD corrections into account, both
\provbfhh and the inclusive stand-alone version of the \provbfhh
program have been extended:
The kappa framework has been implemented in such a way that values for
the coupling modifiers $\kappa_\lambda$, $\kappa_V$, and $\kappa_{2V}$
can be set by the user. In \provbfhh{} the matrix element is expressed
in terms of the DIS structure functions~\cite{Han:1992hr}. In this
formalism the tensor related to the amplitude of the $VV\to HH$
sub-process is expressed as~\cite{Dobrovolskaya:1990kx,Dreyer:2018qbw}
\begin{align}
  \label{eq:VV-subproc}
  \mathcal{M}^{VVHH,\mu\nu} =& 
  2 \sqrt{2} G_F g^{\mu\nu} \bigg(\frac{2{\color{red}\kappa_{V}^2}m_V^4}{(q_1 + k_1)^2 - m_V^2 +i\Gamma_Vm_V}
  + \frac{2{\color{red}\kappa_{V}^2}m_V^4 }{(q_1 + k_2)^2 - m_V^2+i\Gamma_Vm_V}
  \notag\\ &\hspace{2.7cm} + \frac{6 v {\color{red}\kappa_V}{\color{green}\kappa_\lambda} m_V^2}{(k_1 + k_2)^2 - m_H^2+i\Gamma_Hm_H}
  + {\color{blue} \kappa_{2V}}m_V^2 \bigg)
  \notag\\
  &+ \frac{\sqrt{2} G_F {\color{red}\kappa_V^2}m_V^4}{(q_1 + k_1)^2 - m_V^2}
  \frac{(2 k_1^\mu + q_1^\mu)(k_2^\nu - k_1^\nu - q_1^\nu)}{m_V^2-i\Gamma_Vm_V}
  \notag\\ & + \frac{\sqrt{2} G_F {\color{red}\kappa_V^2}m_V^4}{(q_1 + k_2)^2 - m_V^2}
  \frac{(2 k_2^\mu + q_1^\mu)(k_1^\nu - k_2^\nu - q_1^\nu)}{m_V^2-i\Gamma_Vm_V} \,,
\end{align}
where $k_1, k_2$ are the final state Higgs momenta, $q_1$ and $q_2$
are the vector boson momenta and momentum conservation yields $k_1+k_2
= q_1+q_2$. Here $v$ is the vacuum expectation value of the Higgs
field, $G_F$ is Fermi's constant and $\Gamma_V$ and $m_V$ are the
width and mass of the exchanged vector boson, respectively. The
modified couplings are highlighted in colour.

\begin{figure}[pt!]
\centering
\subfloat[][]{\label{fig:qcd-corr-a}
\includegraphics[width=0.4\textwidth]{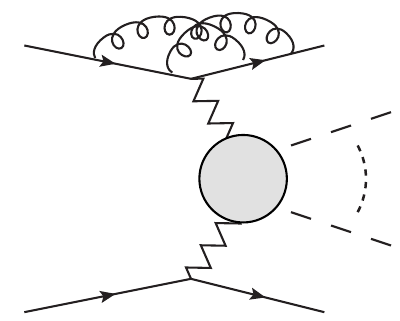}}\hfill
\subfloat[][]{\label{fig:qcd-corr-b}
\includegraphics[width=0.4\textwidth]{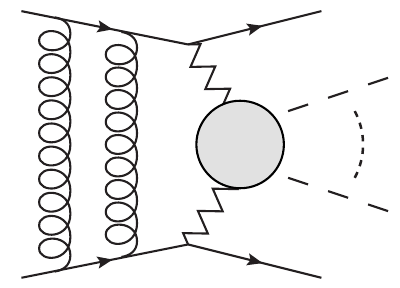}}
\caption{\label{fig:qcd-corr} Representative diagrams for (a)
  factorizable and (b) non-factorizable NNLO-QCD corrections to
  \vbfhh{}. The grey blob represents the $VV\to HH$
  sub-process. Figure reproduced from~\cite{Dreyer:2020urf}.}
\end{figure}
When considering the QCD corrections to the diagrams shown in
Fig.~\ref{fig:diagrams} they can be divided into two separate contributions. The
majority of the corrections come from diagrams where gluons only
attach to a single quark line, cf.\ Fig~\ref{fig:qcd-corr-a}. These corrections are
called \emph{factorized} corrections, as they effectively factorize
from the EW production. Starting at NNLO there are also corrections
stemming from the exhange of two gluons between the two quark lines,
cf.\ Fig.~\ref{fig:qcd-corr-b}. These \emph{non-factorizable} corrections are
colour-suppressed compared to their factorizable counterparts, but
explicit studies of them in the eikonal approximation~\cite{Liu:2019tuy,Dreyer:2020urf,Gates:2023iiv,Asteriadis:2023nyl,Bronnum-Hansen:2023vzh} and
beyond~\cite{Long:2023mvc} have found that they receive a Glauber-phase
enhancement. They are also sensitive to the exact details of the EW
production, since the couplings of the Higgs bosons enter directly in
the loops. In fact, in the Standard Model the non-factorizable
corrections are somewhat suppressed by the unitarity cancellations
between the different diagrams~\cite{Dreyer:2020urf}.

\subsection{Implementation in the \PBOX}
An implementation of the \vbfhh{} process in the framework of the \PBOX{} has not been available until now. We therefore developed such an implementation {from scratch proceeding along similar lines as in the related VBS-induced $\vvjj$ processes~\cite{Jager:2011ms,Jager:2013mu,Jager:2013iza,Jager:2018cyo,Jager:2024sij}.}

{In detail, the implementation of the  \vbfhh{} process required us to prepare matrix elements for all flavour combinations of external partons compatible with charge conservation at each interaction vertex at Born level, virtual and real-emission corrections, as well as colour- and spin correlated amplitudes for the computation of subtraction terms within the FKS-scheme~\cite{Frixione:1995ms} for the treatment of infrared divergences.  To that end, we used the helicity-amplitude formalism of Ref.~\cite{Hagiwara:1988pp} for the computation of fermionic currents, amended by building blocks for bosonic tensors accounting for the $VV\to HH$ subprocess within the SM and the kappa framework prepared with the help of a heavily customized version of the {\tt Madgraph~\!2} amplitude generator~\cite{Stelzer:1994ta}.   
At Born level, only subprocesses of the type $q q'\to q q' HH$ (where
$q$ and $q'$ denote any light quark flavour) and crossing-related ones
with external anti-quarks appear. The virtual contributions comprise
gluonic loop corrections to either the upper or the lower quark line,
resulting in a total expression proportional to the Born amplitude,
similar to the case of VBF-induced single Higgs production discussed
in \cite{Nason:2009ai}, such that the finite part of the virtual
corrections, $\vfin$, assumes the form
\beq \vfin = C_F \left[
  -\ln^2\left(\frac{\mur^2}{q_1^2}\right)
  -3\ln\left(\frac{\mur^2}{q_1^2}\right)
  -\ln^2\left(\frac{\mur^2}{q_2^2}\right)
  -3\ln\left(\frac{\mur^2}{q_2^2}\right) -16 \right] \mc{B}\,, \eeq As
above, $q_1$ and $q_2$ denote the momenta of the two bosons exchanged
in the $t$-channel, $\mur$ is the renormalization scale, $C_F=4/3$,
and $\mc{B}$ the Born amplitude.

For the real-emission contributions, matrix elements for all
subprocesses with an additional external gluon $g$ had to be
developed. Similar to the Born case, for the computation of all
subprocesses of the type $q q'\to q q' gHH$ and crossing-related
contributions we used the helicity-amplitude formalism of
Ref.~\cite{Hagiwara:1988pp} supplemented by customized bosonic tensors
for the $VV\to HH$ sub-amplitudes.
All matrix elements were implemented in such a way that, depending on
the parameters chosen by the user in an editable input file, they can
account for the \vbfhh{} production process in the SM, but also in the
kappa framework with adjustable values of the couplings modifiers
$\klam, \khhvv,\khvv$.


In addition to the partonic scattering amplitudes, the implementation
of the \vbfhh{} process in the \POWHEGBOXVV{} required us to provide a
customized phase-parameterization.  To that end, we adapted a
respective routine of the \provbfhh program to comply with the format
requirements of the \PBOX.
Since the inclusive leading-order (LO) cross section for \vbfhh{} is
finite, no technical cuts are needed to obtain numerically stable
results.  }

In order to validate our implementation, we have compared tree-level
matrix elements for selected phase-space points with
\MADGRAPH-generated ones, finding full agreement. We have checked that
the real-emission contributions approach their soft and collinear
limits correctly. Results for a variety of kinematic distributions of
all final-state particles at LO and NLO-QCD have been compared to
those obtained with the \provbfhh program both within the SM and for
non-unit values of the coupling modifiers. We found full agreement in
each case.

\section{Phenomenological results}
\label{sec:pheno}
In this section we present some representative phenomenological
results using our new implementations.
\subsection{Setup}
We consider proton-proton collisions at the LHC with a centre-of-mass
energy of $\sqrt{s} = 13 \, \mathrm{TeV}$.  For the parton shower
results we use \texttt{PYTHIA}
version~8.312~\cite{bierlich2022comprehensiveguidephysicsusage} with
the \texttt{Monash 2013} tune~\cite{Skands_2014} and \texttt{HERWIG7} version~7.3.0~\cite{Bewick:2023tfi}. We have adapted the matching procedure for \HERWIGS{} and \PYTHIA{} from Ref.~\cite{FerrarioRavasio:2023jck} and Ref.~\cite{Banfi:2023mhz}, i.e.\ we use the options
\begin{flalign*}
&{\tt set /Herwig/Shower/ShowerHandler:MaxPtIsMuF Yes}&\\ 
&{\tt set /Herwig/Shower/ShowerHandler:RestrictPhasespace Yes}&
\end{flalign*}
in \HERWIGS{} and the settings
\begin{flalign*}
&{\tt POWHEG:veto = 1}&\\
&{\tt POWHEG:pThard = 0}&\\
&{\tt POWHEG:pTemt = 0}&\\
&{\tt POWHEG:emitted = 0}&\\
&{\tt POWHEG:pTdef =1}&
\end{flalign*}
in the context of the \texttt{PowhegHooks} class contained in \PYTHIA.

Both the
dipole-local \texttt{PYTHIA} shower~\cite{Cabouat:2017rzi} and the
\VINCIA{} antenna shower~\cite{Fischer:2016vfv} are
considered. As has been noted in the past the default \texttt{PYTHIA}
shower should be avoided, in particular for VBF processes, as it
breaks coherence which leads to an excess of central jet
activity~\cite{Ballestrero:2018anz,Jager:2020hkz,Hoche:2021mkv}. Underlying
event (UE), hadronization, multi-parton interactions (MPI), and QED radiation effects are turned off.

For the parton distribution functions (PDFs) of the protons we use the PDF set \texttt{NNPDF40MC\_nnlo\_as\_01180}~\cite{cruzmartinez2024lonlonnloparton} as obtained from the {\tt LHAPDF6} repository~\cite{Buckley:2014ana} with the associated strong coupling $\alpha_s(m_Z) = 0.118$. The number of active quark flavours is set to five. 
For the values of the masses and widths of the $W$, $Z$ and $H$ bosons we use~\cite{ParticleDataGroup:2024cfk}:  
\begin{align}\label{masses_and_widths}
m_W &= 80.3692 \, \mathrm{GeV}, \qquad \Gamma_W = 2.085 \, \mathrm{GeV}\,,\\
m_Z &= 91.1880 \, \mathrm{GeV}, \qquad \Gamma_Z = 2.4955 \, \mathrm{GeV}\,,\\
m_H &= 125.2 \, \mathrm{GeV}, \qquad \Gamma_H = 3.7 \times 10^{-3} \, \mathrm{GeV}\,.
\end{align}
We apply the $G_{\mu}$~scheme~\cite{Denner:2000bj}, where $\alpha$ and
the weak mixing angle are calculated from the Fermi constant $G_{\mu}
= 1.16637~\times~10^{-5}~\mathrm{GeV}^{-2}$, $m_W$ and $m_Z$ via
tree-level EW relations. For the Cabibbo-Kobayashi-Maskawa matrix a
diagonal form is assumed and we apply the \texttt{Bornzerodamp}
mechanism as already implemented in the \PBOX{}, to dynamically
separate the real emission matrix element into its singular and
non-singular part. Since the negative-weight fraction is not
negligible (about $20\%$ in a standard run), we also make use of
folding~\cite{Nason:2007vt} to reduce the fraction to a few percent.

The factorization and renormalization scales, $\mur=\xir\mu_0$ and $\muf=\xif\mu_0$, are set according to
\begin{equation}\label{scale}
\mu_0^2 =\dfrac{m_H}{2}\sqrt{\left(\frac{m_H}{2}\right)^2+p^2_{T,HH}}\,, 
\end{equation}
where $p_{T,HH}$ denotes the transverse momentum of the Higgs-pair system. 
Scale uncertainties are estimated by a 7-point variation of the scale factors $\xir$ and $\xif$ by factors between 0.5 and 2.

For our numerical studies we reconstruct jets according to the anti-$k_T$ algorithm~\cite{Cacciari:2008gp} with a distance parameter of $R=0.4$ using the {\tt FastJet} package~\cite{Cacciari:2011ma}, version 3.3.4.  
We require at least two hard jets $j$ with transverse momenta and rapidities in the range  
\begin{equation}
\label{eq:cuts1}
p_{T,j} > 25 \, \mathrm{GeV}, \qquad \vert y_j \vert < 4.5, 
\end{equation}
The two hardest jets fulfilling this criterion are referred to as {\em tagging jets}. The two tagging jets are required to exhibit a large invariant mass and rapidity separation, 
\begin{equation}
\label{eq:cuts2}
\qquad \mjjtag > 600 \, \mathrm{GeV}\,,\qquad
 \Delta \yjjtag = |y_{j_1}^\mr{tag}-y_{j_2}^\mr{tag}|> 4.5, \qquad 
 y_{j_1}^\mr{tag} \cdot y_{j_2}^\mr{tag} < 0.
\end{equation}
Whenever we consider subleading jets, we apply additional cuts to make these jets well-identifiable. For instance, for some distributions of the third-hardest jet, to be discussed below, in addition to the cuts of Eqs.~\eqref{eq:cuts1}--\eqref{eq:cuts2}, we require 
\begin{equation}
\label{eq:cuts3}
p_{T,j_3} > 25 \, \mathrm{GeV}, \qquad \vert y_{j_3} \vert < 4.5.
\end{equation} 
No cuts are applied to the Higgs bosons. 

The results for the fiducial cross section after cuts at different
accuracy are shown in Tab.~\ref{table1}.

\begin{table}[h]
\begin{center}
\begin{tabular}[t]{lcc}
\toprule
accuracy & $\sigma$ [fb]  & ratio to NLO    \\
\midrule
LO              &   $ 0.668   $ & $1.087$   \\
NLO             &   $ 0.614  $ &  $1$       \\
NNLO            &   $ 0.603  $ & $ 0.982 $  \\
\midrule
NLO+\PYTHIAE        &   $ 0.585^{+0.007}_{-0.009}  $ & $0.953 $  \\
NLO+\VINCIA         &   $ 0.592^{+0.007}_{-0.010}  $ &  $0.964$ \\
NLO+\HERWIGS        &   $ 0.575^{+0.013}_{-0.007}  $ & $0.936$ \\
\bottomrule
\end{tabular}
\caption{\label{table1} SM fiducial cross sections in fb after the cuts of Eqs.~\eqref{eq:cuts1}--\eqref{eq:cuts2} at different accuracies and ratio to NLO. Statistical errors are beyond the quoted precision. Scale uncertainties are indicated by the subscript and superscript for NLO+PS accuracy only.}
\end{center}
\end{table}

\subsection{Parton-shower matched results}
\begin{figure}[pt!]
\centering
\subfloat[][]{
\includegraphics[width=0.5\textwidth]{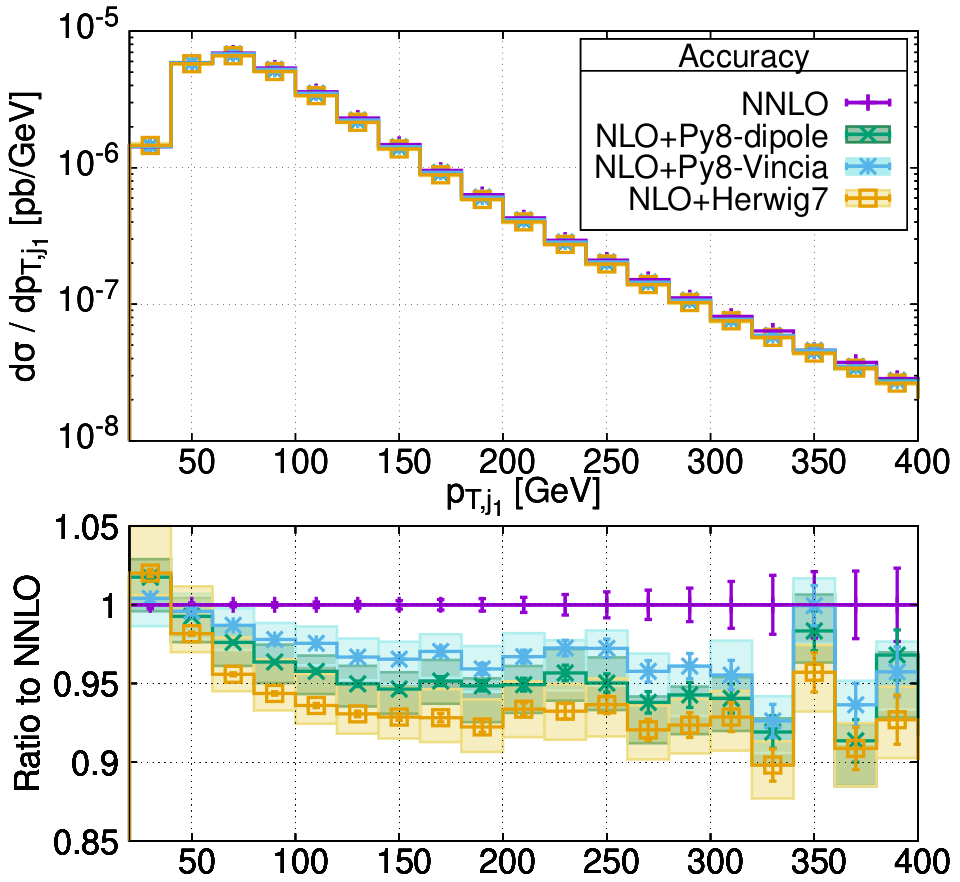}}
\subfloat[][]{
\includegraphics[width=0.5\textwidth]{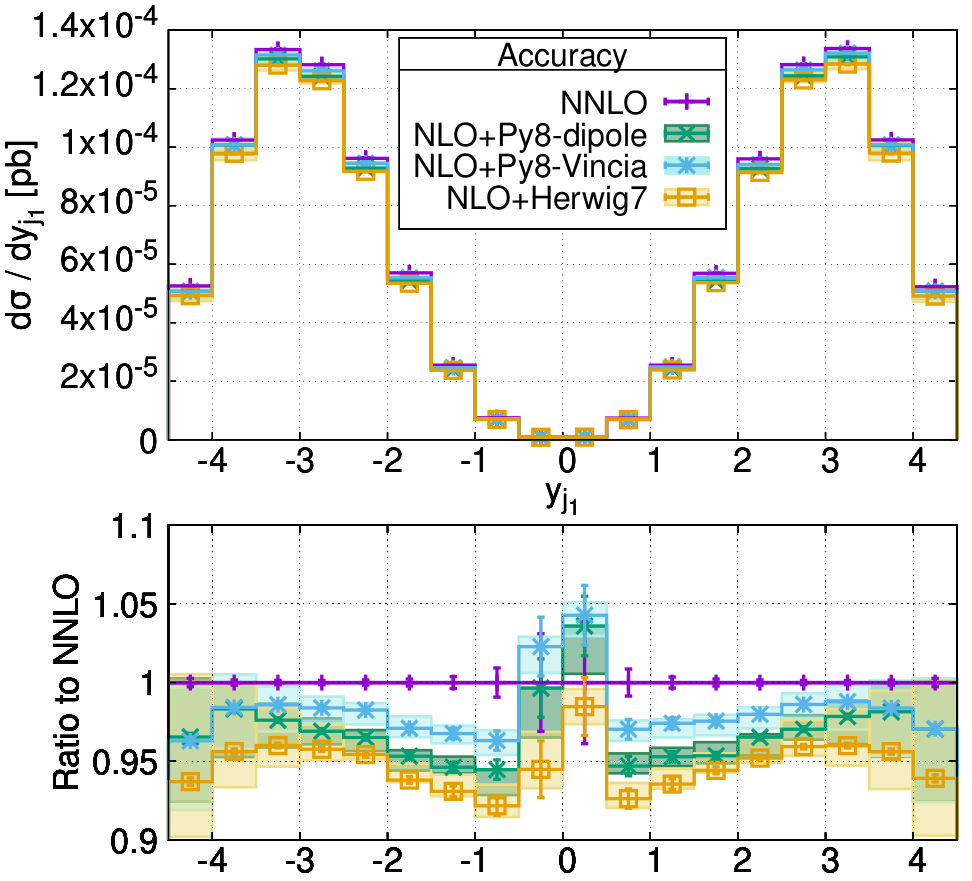}}
\caption{\label{fig:jet1-sm} 
Transverse-momentum (left) and rapidity distributions (right) of the hardest tagging jet for the \vbfhh{} process as described in the text within the cuts of Eqs.~\eqref{eq:cuts1}--\eqref{eq:cuts2} at NNLO (magenta), NLO+PS using \HERWIGS{} (orange) or  \PYTHIAE{} with the dipole shower (green) and the \VINCIA{}  shower (blue), and their ratios to the respective NNLO results (lower panels). Error bars indicate statistical uncertainties, bands correspond to a 7-point variation around the central scale $\mu_0$ of each curve. } 
\end{figure}

Let us first consider \vbfhh{} production in the framework of the SM. In Fig.~\ref{fig:jet1-sm} 
the transverse-momentum and rapidity distributions of the hardest
tagging jet are shown for the NNLO and the NLO+PS predictions using
either \HERWIGS{} or \PYTHIAE{}, for the latter using either the
``dipole'' shower or the \VINCIA{} shower.
We find relatively good agreement between all predictions with a
slight redistribution of events from high to low values of transverse
momentum in the NLO+PS results as compared to the NNLO predictions,
and a moderate change of shape in the rapidity distributions, with the
NLO+PS results tending to more central values.  The different shower
options yield similar results, with the \VINCIA{} shower being closest
to the NNLO benchmark results and the \HERWIGS{} shower being furthest
away from the NNLO values. However, the differences between
  the various showers are mostly contained in the normalization, with
  very similar shapes across the three variants.

\begin{figure}[pt!]
\centering
\subfloat[][]{
\includegraphics[width=0.5\textwidth]{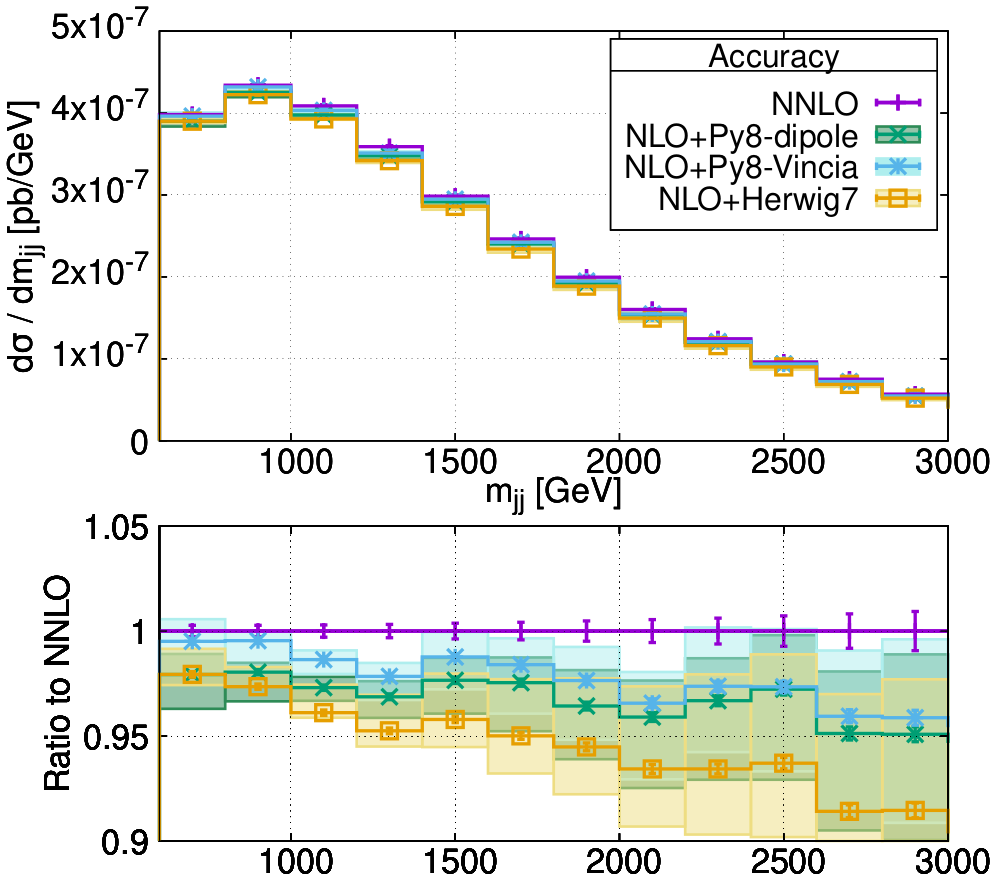}}
\subfloat[][]{
\includegraphics[width=0.5\textwidth]{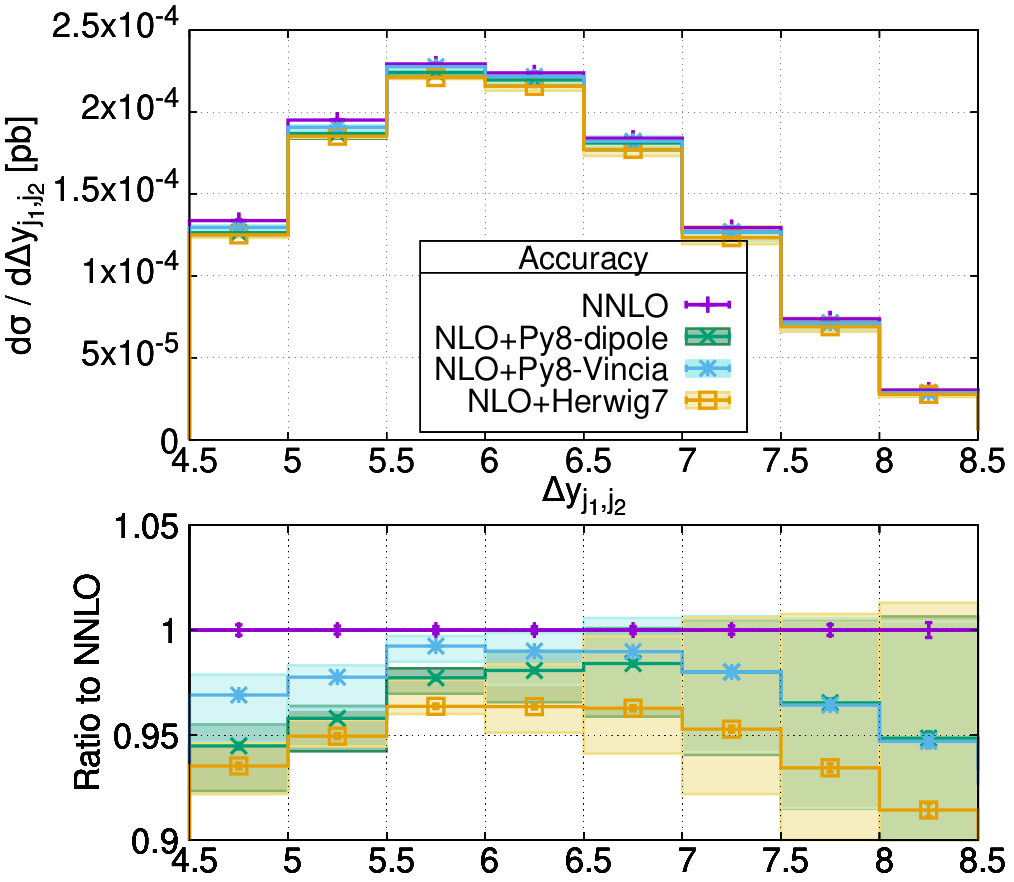}}
\caption{\label{fig:mjj-yjj-sm} 
Similar to Fig.~\ref{fig:jet1-sm}, but for the invariant mass distribution (left) and the rapidity separation (right) of the two tagging jets.  }
\end{figure}

A similar tendency can be observed in invariant mass distribution and rapidity separation of the two tagging jets, shown in Fig.~\ref{fig:mjj-yjj-sm}. 
For these distributions the NLO+PS predictions are consistently below the fixed-order predictions. In particular, at large values of $\mjjtag$ the extra radiation emerging because of the PS results in a slight shift of the taggings jets' kinematics, resulting in fewer events passing the selection cuts. This feature is rather genuine and does not change if instead of \HERWIG{} either of the \PYTHIAE{} shower options is used.

\begin{figure}[pt!]
\centering
\subfloat[][]{
\includegraphics[width=0.5\textwidth]{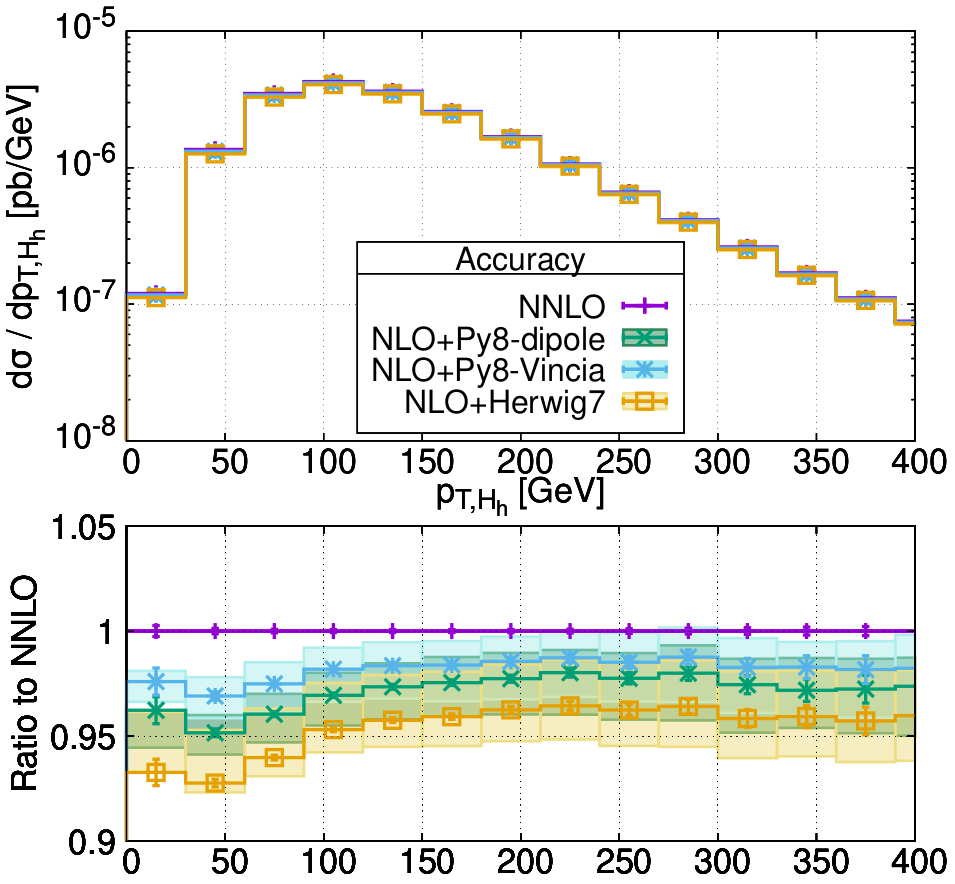}}
\subfloat[][]{
\includegraphics[width=0.5\textwidth]{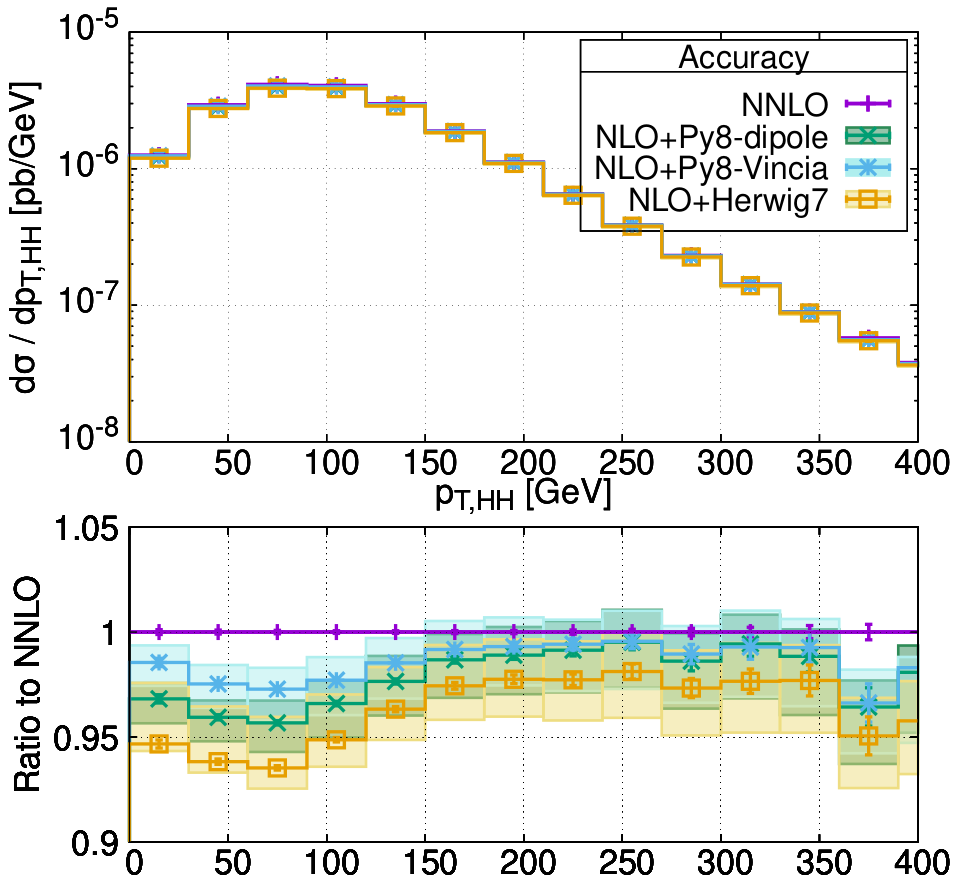}}
\caption{\label{fig:pth-pthh-sm} 
Similar to Fig.~\ref{fig:jet1-sm}, but for the transverse momentum distribution of the hardest Higgs boson (left) and of the Higgs-pair system (right).  }
\end{figure}
In Fig.~\ref{fig:pth-pthh-sm} the transverse momentum distributions of the hardest Higgs boson and the Higgs-pair system are shown. These exhibit a similar trend as the distributions of the transverse momentum of the hardest tagging jet and of the invariant mass of the tagging jet system with the NLO+PS predictions lying a few percent below the fixed-order NNLO predictions.

\begin{figure}[pt!]
\centering
\subfloat[][]{
\includegraphics[width=0.5\textwidth]{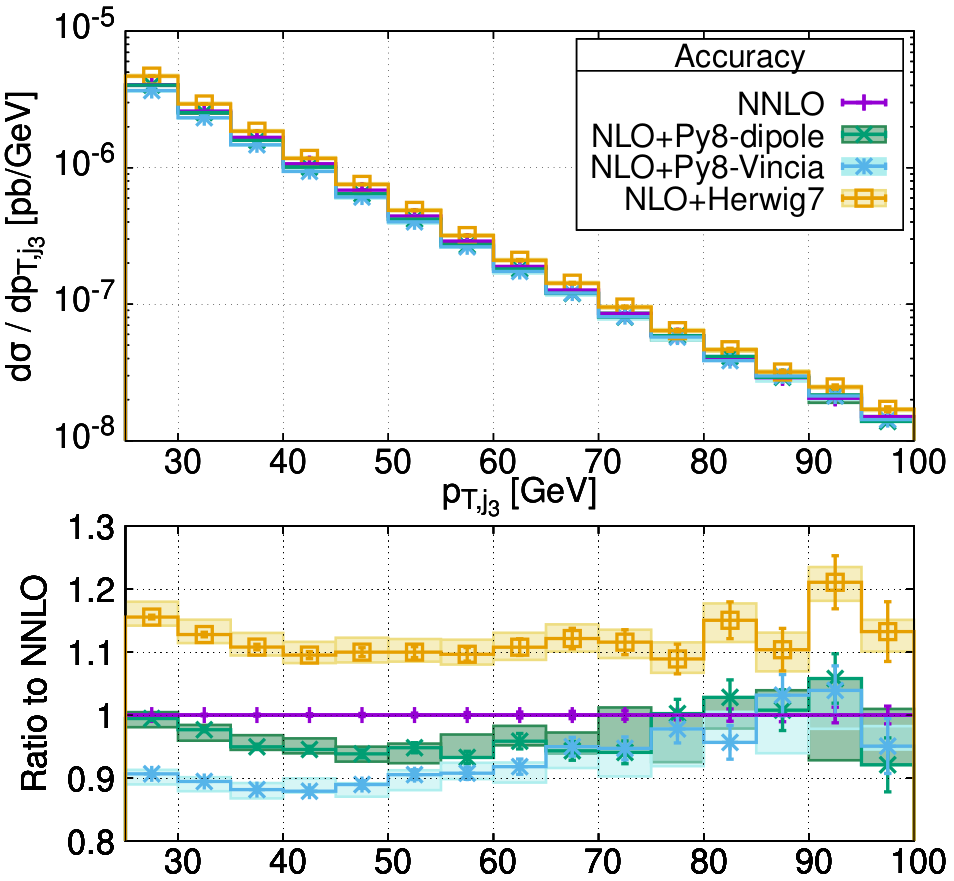}}
\subfloat[][]{
\includegraphics[width=0.5\textwidth]{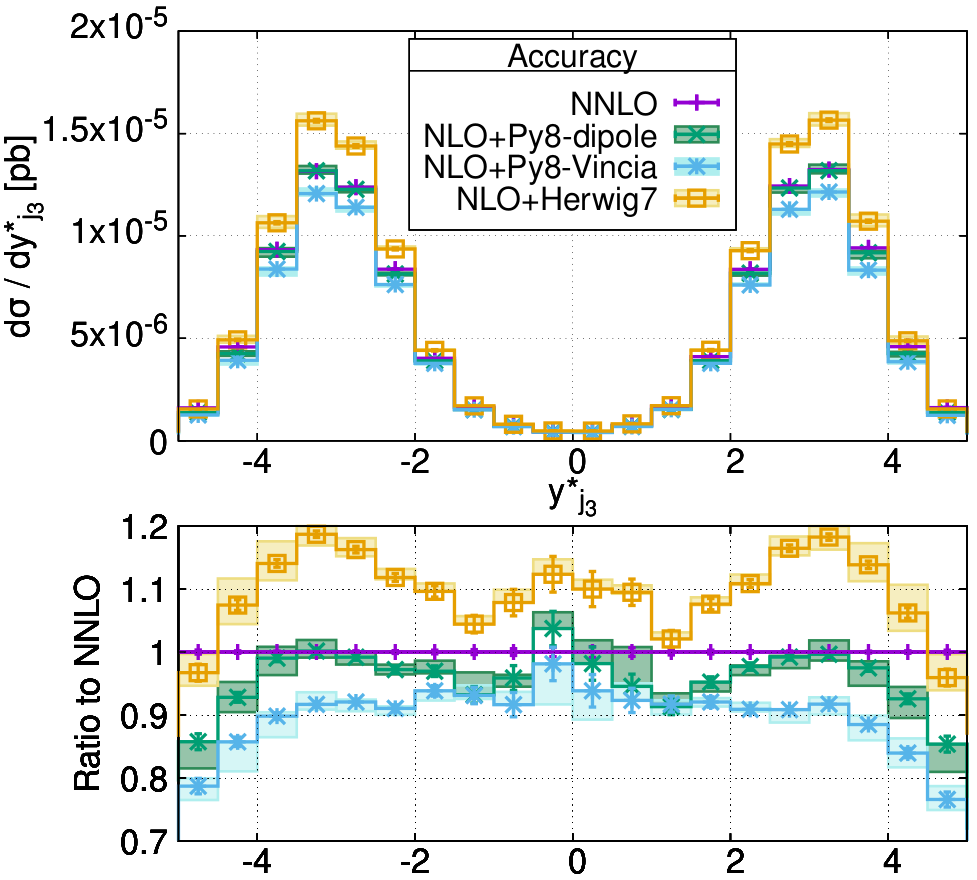}}
\caption{\label{fig:jet3-sm} 
Similar to Fig.~\ref{fig:jet1-sm}, but for the transverse-momentum and the relative rapidity distribution of the third jet as defined in Eq.~\eqref{eq:ystar}. The additional cuts of Eq.~\eqref{eq:cuts3} are applied.  }
\end{figure}

In Fig.~\ref{fig:jet3-sm}
%
the transverse-momentum and relative rapidity of the third jet with respect to the tagging-jet system are considered. The latter is defined as 
\beq
\label{eq:ystar}
\yjstar = y_{j_3}- \frac{y_{j_1}+y_{j_2}}{2}\,. 
\eeq
We note that a third jet only enters via the real-emission contributions in the NLO calculation of the \vbfhh{} process, and it thus only accounted for with tree-level accuracy in our NLO results. NLO accuracy for distributions of the third jet is achieved by the NNLO calculation of the $\hhjj$ process. In the NLO+PS predictions, a third jet can also result from PS emissions. 
As apparent from Fig.~\ref{fig:jet3-sm}, the NLO+PS prediction using
the dipole version of \PYTHIAE{} provides a decent approximation of
the NNLO results, whereas \VINCIA{} sits somewhat below.  The
\HERWIGS{} predictions, on the other hand, overshoot the \PYTHIAE{}
results by 10~to~20\%. The larger spread in predictions is expected
due to the lower perturbative accuracy for the third jet, but should
be taken into account whenever the third jet enters an experimental
analysis. In order to reduce this uncertainty, the third jet would
have to be matched at NLO, as was done in single Higgs VBF
production~\cite{Jager:2014vna}.

\subsection{Hadronization and underlying event}
\begin{figure}[pt!]
\centering
\subfloat[][]{
\includegraphics[width=0.5\textwidth]{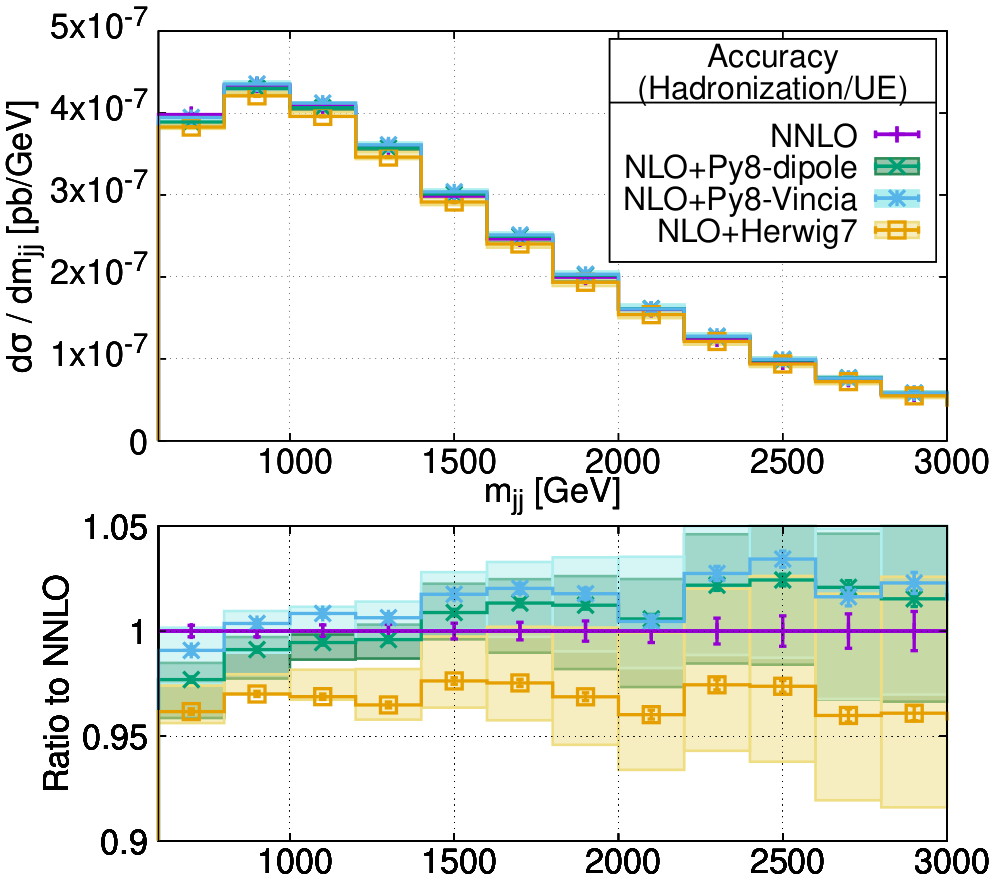}}
\subfloat[][]{
\includegraphics[width=0.5\textwidth]{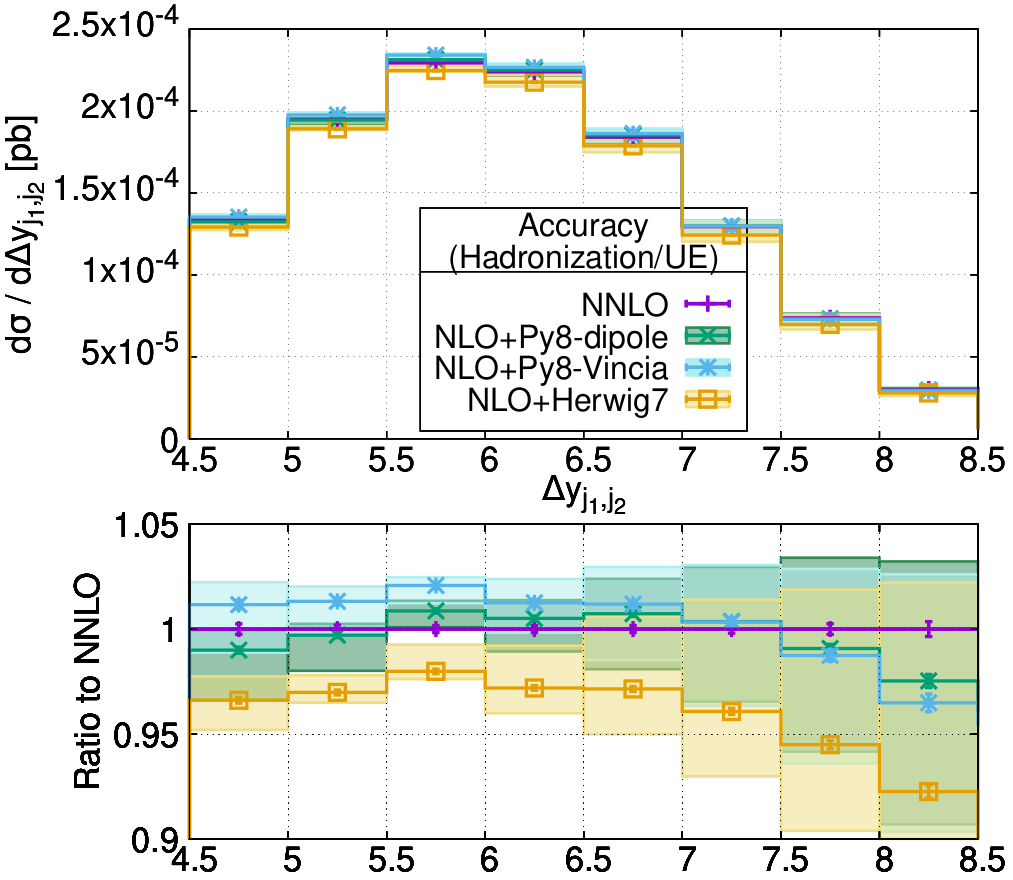}}
\caption{\label{fig:hadronization} 
Similar to Fig.~\ref{fig:mjj-yjj-sm}, but with underlying event and hadronization turned on.}
\end{figure}

\begin{figure}[pt!]
\centering
\subfloat[][]{
\includegraphics[width=0.5\textwidth]{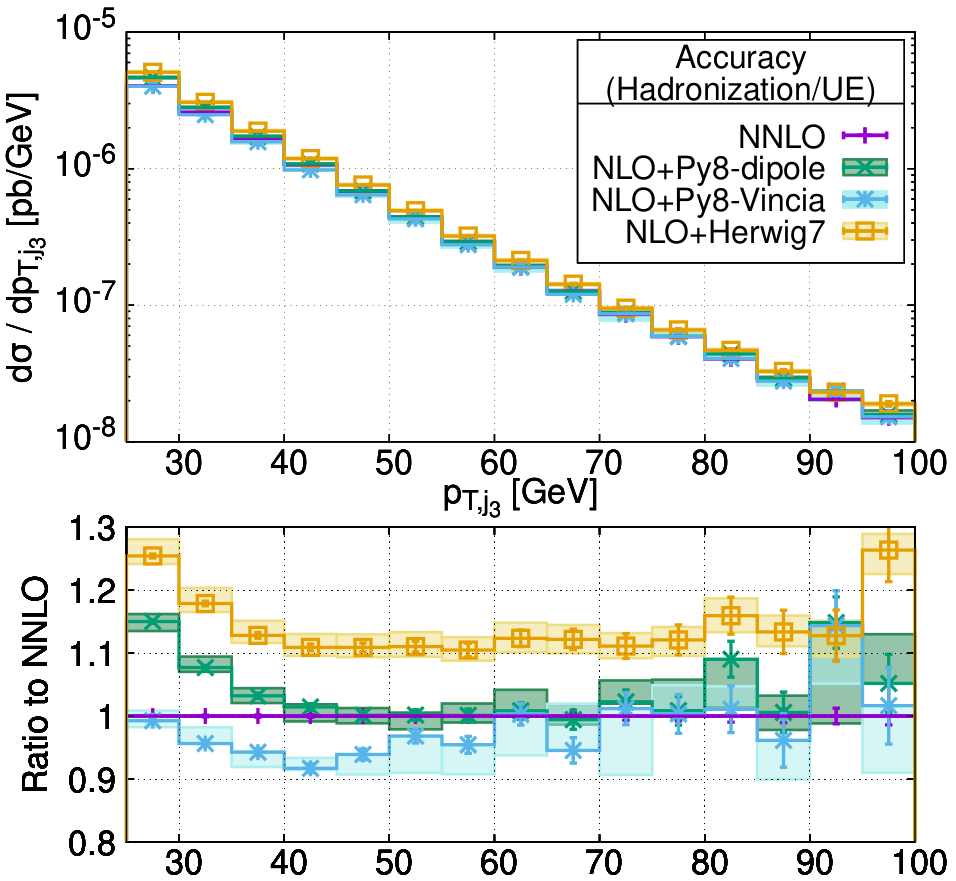}}
\subfloat[][]{
\includegraphics[width=0.5\textwidth]{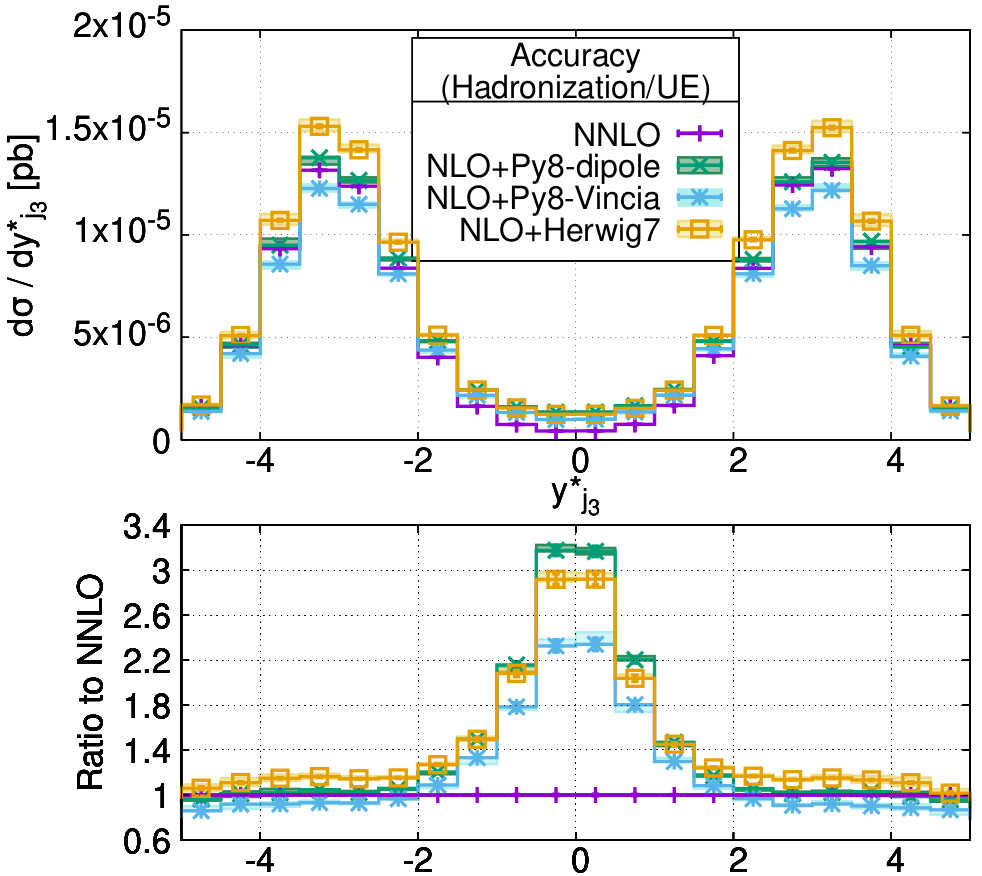}}
\caption{\label{fig:hadronization2} 
Similar to Fig.~\ref{fig:jet3-sm}, but with underlying event and hadronization turned on.}
\end{figure}

To further study the impact of parton shower settings on the
observables we also display the results with hadronization and
underlying event turned on in the parton shower in
Fig.~\ref{fig:hadronization} for the invariant mass distribution and
the rapidity separation of the two tagging jets, and in
Fig.~\ref{fig:hadronization2} for the transverse momentum and the
relative rapidity distribution of the third jet.

We observe that for observables concerning the tagging jets
  the hadronization/UE only has a small impact increasing the ratio to
  NNLO by a few percent while keeping the relative spread of the
  NLO+PS predictions mostly unchanged. The impact on the third jet
  observables is much larger filling the central region between the
  two tagging jets as apparent from Fig.~\ref{fig:hadronization2}
  (right). This effect has been observed in the past for other VBF
  processes~\cite{Bittrich:2021ztq}. 

  \subsection{Impact of anomalous couplings}
With these features of the perturbative corrections for the SM predictions in mind, let us now turn towards  an analysis of anomalous Higgs couplings using the kappa framework introduced in Sec.~\ref{sec:implementation}. Since the NLO+PS predictions have been found to provide a good approximation of the full NNLO results, in the following we will only show results obtained with our \PBOX{} implementation using the \VINCIA{} shower~\cite{Fischer:2016vfv} of \PYTHIAE{}.

We consider values of the coupling factors compatible with experimental bounds, i.e.\ 
$0.6< \khhvv <1.5$~\cite{ATLAS:2024ish}, 
$0.98< \khvv <1.1$~\cite{CMS:2024awa} (we have extracted the $2\sigma$ limit from Fig.~6), 
$-1.2< \klam <7.2$~\cite{ATLAS:2024ish}.

In Tab.~\ref{table2} we display the fidcuial cross section results
after VBF cuts for different values of the anomalous couplings and the
ratio to the SM values at NLO+PS, using the \VINCIA{} shower
prediction.

\begin{table}[b!]
\begin{center}
\begin{tabular}{lcc}
\toprule
Coupling factor & $\sigma$ [fb]  & ratio to SM    \\
\midrule
SM                        &   $ 0.592 $ & $1$   \\
\midrule
$\kappa_{2V}=1.5 $        &   $ 1.14  $ & $1.93$  \\
$\kappa_{2V}=0.6 $        &   $ 2.57  $ & $ 4.34$  \\
$\kappa_{V}=1.1$          &   $ 1.86  $ & $3.15$  \\
$\kappa_{V}=0.98$         &   $ 0.463 $ & $0.781$ \\
$\kappa_{\lambda}=7.2$    &   $ 13.0  $ & $21.9$ \\
$\kappa_{\lambda}=-1.2$   &   $ 3.71  $ & $6.27$ \\
\bottomrule
\end{tabular}
\caption{\label{table2} Fiducial cross sections in fb after the cuts
  of Eqs.~\eqref{eq:cuts1}--\eqref{eq:cuts2} for the different
  anomalous couplings and the ratio to the SM value for
  NLO+\VINCIA{}. Statistical errors are beyond the quoted precision.}
\end{center}
\end{table}

\begin{figure}[pt!]
\centering
\subfloat[][]{
\includegraphics[width=0.5\textwidth]{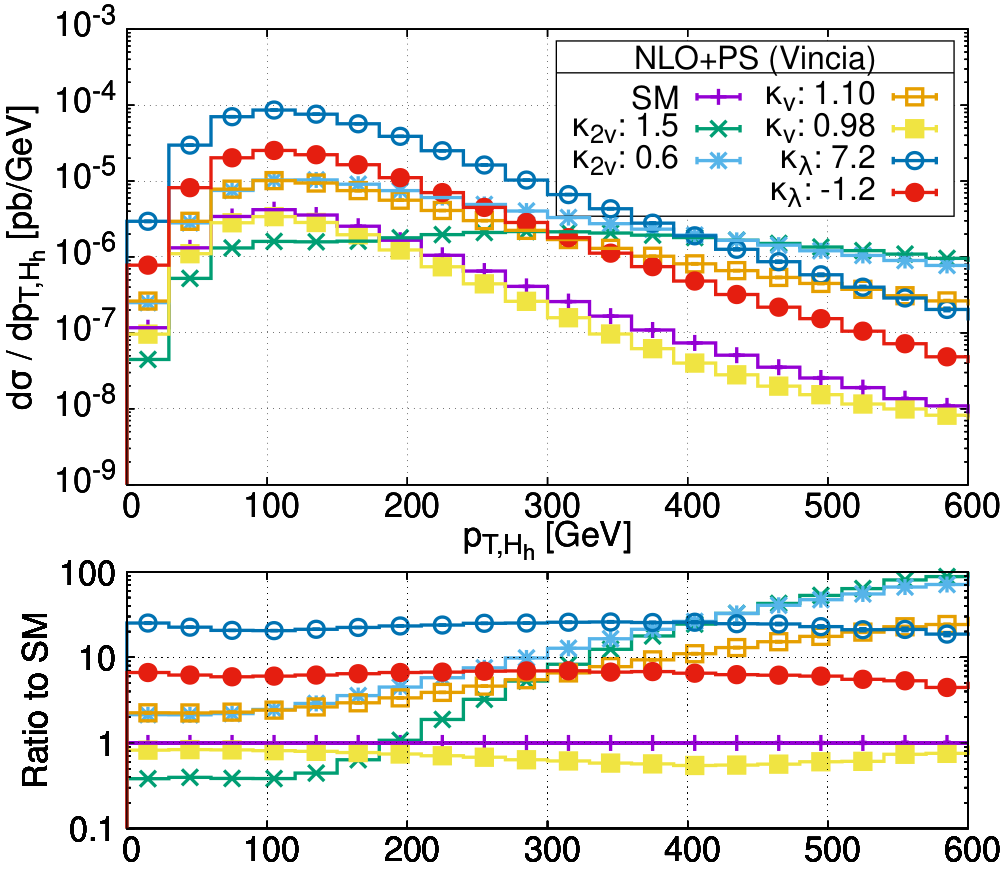}}
\subfloat[][]{
\includegraphics[width=0.5\textwidth]{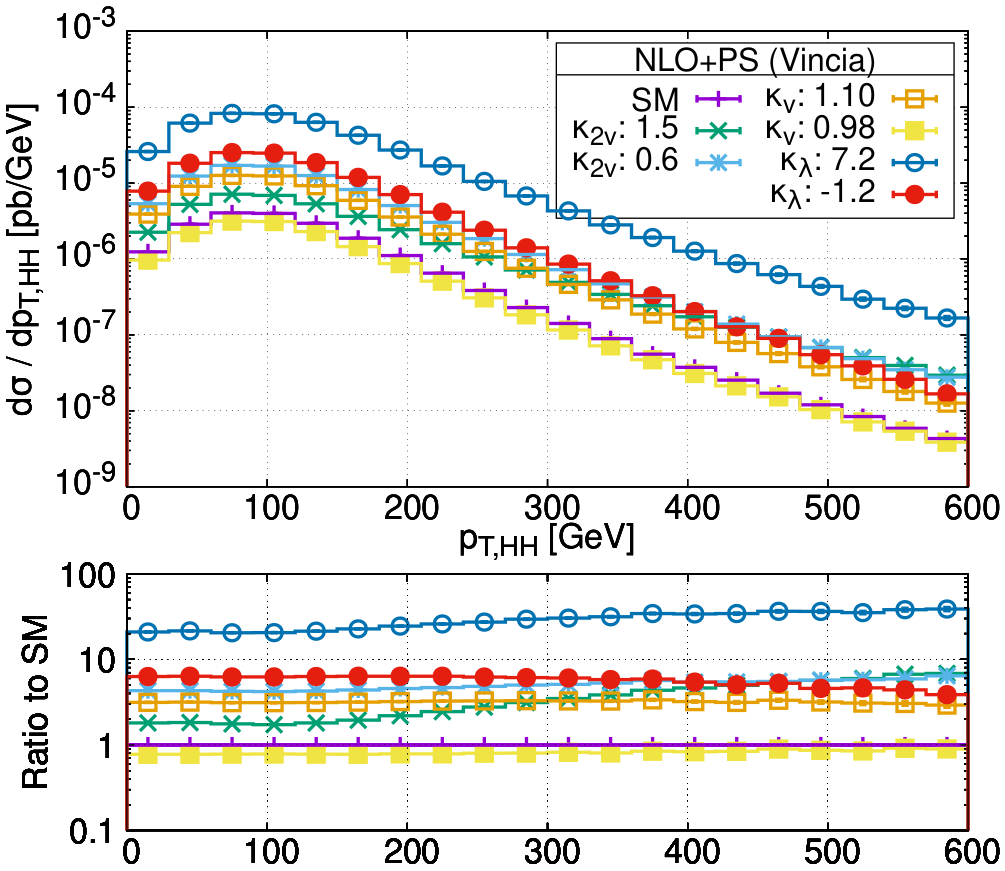}}
\caption{\label{fig:pth-kappa} Transverse-momentum distribution of the
  hardest Higgs boson (left) and of the Higgs-pair system (right) for
  the \vbfhh{} process as described in the text within the cuts of
  Eqs.~\eqref{eq:cuts1}--\eqref{eq:cuts2} at NLO+PS accuracy within
  the SM (magenta), and for different values of the Higgs coupling
  modifiers $\kappa_i$ as indicated in the legend, together with their
  ratios to the respective SM results. Statistical uncertainties are
  indicated by error bars (but mostly too small to be visible).}
\end{figure}

In Fig.~\ref{fig:pth-kappa} 
the transverse-momentum distributions of the hardest Higgs boson and
of the Higgs-pair system are considered within the SM, and for
selected scenarios with one coupling modifier being set to a non-SM
value compatible with experimental bounds while all other couplings
retain their SM value. When the coupling modifiers are set to values
significantly different from $1$, both of the considered distributions
change their normalization and shape considerably. This is due
  to the very delicate unitarity cancellations which are present in
  the Higgs sector. The large sensitivity to these couplings highlights 
  the importance of the \vbfhh{} process in the Higgs program. 
It is clear that for such large deviations of the couplings from their
SM values the validity of the approach breaks down. More refined
models would be needed to understand the very mechanism resulting in a
genuinely non-SM type behavior of the relevant Higgs
couplings. Indeed, the kappa framework must be applied only to {\em
  identify} deviations from the SM expectation, but cannot be expected
to provide a deeper understanding of the origin of such
deviations. For instance, the so-called $K$-matrix unitarisation
scheme to regulate the high-energy behavior of VBF and vector boson
scattering processes in the presence of physics beyond the SM has been
explored in~\cite{Alboteanu:2008my}.
We believe that nonetheless it is useful to provide tools capable to
compute predictions in a framework that is used by many experimental
analyses to simplify the comparison of data with theoretical
predictions.

Finally, we note that the interplay between the anomalous couplings
and the NLO-QCD corrections are expected to be very mild, as the
corrections completely factorize from the EW production of the two
Higgs bosons. We have verified that the NLO corrections to the
fiducial cross section in this setup vary by less than a percent. This
can also be observed for the total inclusive cross section in figures
8 and 9 in Ref.~\cite{Dreyer:2018qbw}.

\subsubsection{Inclusive results at N$^3$LO}
Our discussion so far has been focused on the new \PBOX{}
implementation. As mentioned already in the introduction we have also
implemented the anomalous couplings in the \provbfhh program. That
implementation served as a cross check of the \PBOX{} implementation,
but can in particular also be used to provide high accuracy
predictions for quantities inclusive in the jet kinematics, such as
the inclusive cross section and Higgs boson transverse momentum.

\begin{figure}[t!]
\centering
\includegraphics[width=0.9\textwidth]{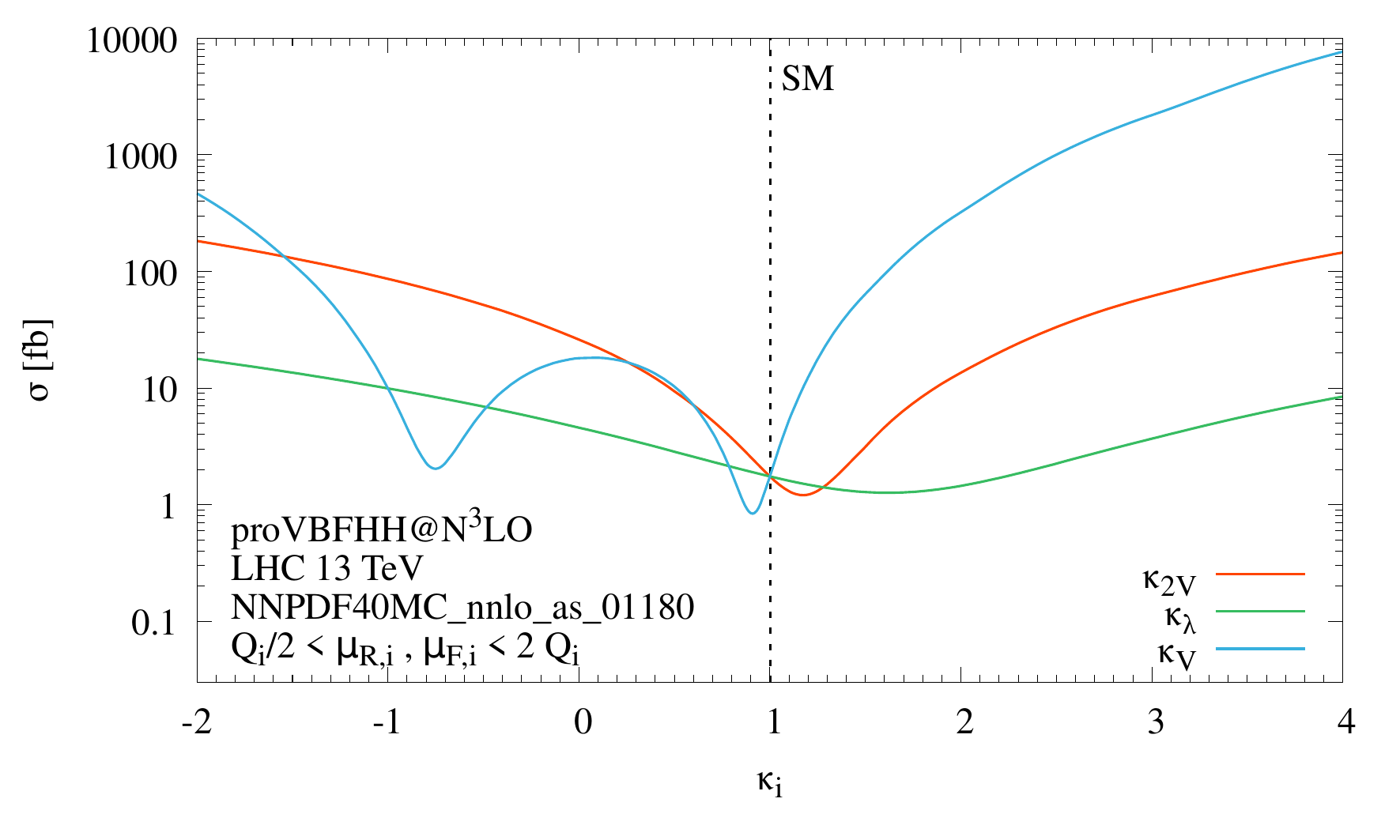}
\caption{\label{fig:coupling-scan} The inclusive \vbfhh{} cross
  section in fb as a function of $\kappa_{2V}$ (red), $\kappa_\lambda$
  (green), and $\kappa_V$ (blue) using \provbfhh{} at N$^3$LO. The SM
  value is indicated with a dashed line. The QCD scale uncertainty is
  entirely contained within the line width of the plot.
}
\end{figure}
To showcase this, in Fig.~\ref{fig:coupling-scan} we show a scan of
the three coupling modifiers $\kappa_{2V}$, $\kappa_\lambda$, and
$\kappa_V$ in the range $[-2:4]$ in the same setup as above at
N$^3$LO, but without any cuts imposed. The renormalization and
factorization scales are chosen separately for each quark line, as the
momentum squared of the vector boson attached to that quark,
$-q_i^2$. The scales are varied by a factor two up and down, yielding
a variation in the permille range, which is completely contained
within the line width in the plot. The figure clearly shows the very
strong dependence of the cross section on the coupling modifiers. From
this plot it is also clear that one cannot fully distinguish the three
couplings from an inclusive measurement alone, but rather that one
needs to complement the inclusive information with distributions
obtained with for instance our new \PBOX{} implementation. In fact,
there must exist a hypersurface in the $\{\kappa_{2V}, \kappa_\lambda,
\kappa_V \}$-space where the inclusive cross section is identical to
the SM value, making an inclusive measurement in that case
useless. One advantage, however, of the \provbfhh{} code, beyond its
state-of-the-art perturbative accuracy, is that it is extremely fast,
and that plots like the above scan can in principle be obtained on a
laptop.

We note that the more complicated dependence on $\kappa_V$ compared to
the two other coupling modifiers comes down to the fact that this
coupling enters quadratically in Eq.~\eqref{eq:VV-subproc} as opposed
to linearly. At the level of the cross section this means that the
coupling enters quartically as opposed to quadratically. As stated
already above, the factorized QCD corrections are almost constant with
respect to the anomalous couplings.

\subsubsection{Impact on non-factorizable corrections}
\label{sec:impact-non-fact}
\begin{figure}[t!]
\centering
\includegraphics[width=0.9\textwidth]{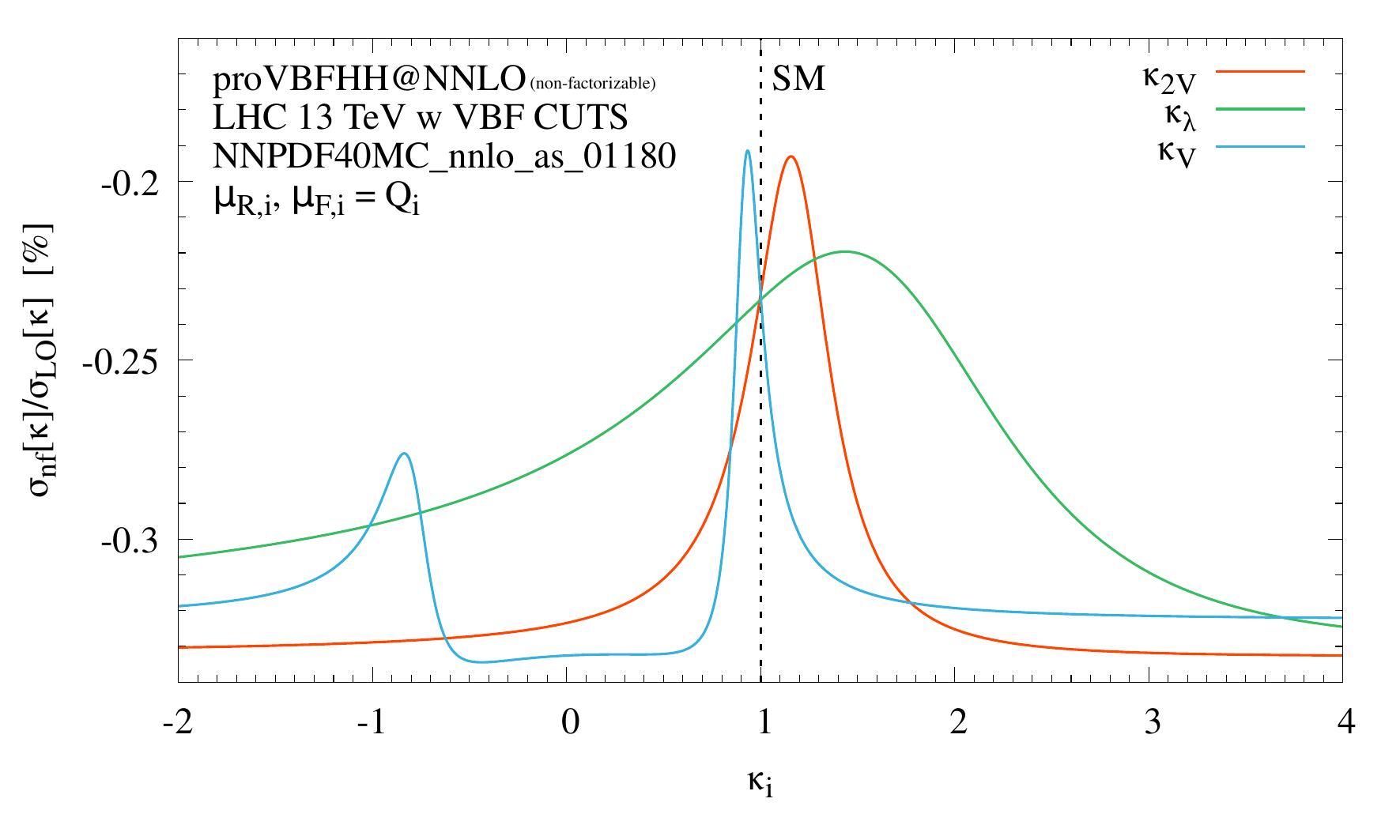}
\caption{\label{fig:coupling-scan-nonfact} The relative impact of
  non-factorizable NNLO corrections to the \vbfhh{} process in percent
  as a function of $\kappa_{2V}$ (red), $\kappa_\lambda$ (green), and
  $\kappa_V$ (blue) obtained with \provbfhh{}. The SM value is
  indicated with a dashed line. The cross section is computed within
  the VBF cuts of Eqs.~\eqref{eq:cuts1}--\eqref{eq:cuts2}.}
\end{figure}
Unlike the factorizable corrections discussed above the
non-factorizable corrections are very sensitive to the anomalous
couplings. This can be understood by realising that the anomalous
couplings spoil the delicate unitarity cancellation taking place in
the $VV \to HH$ sub-process. Since the non-factorizable corrections
act slightly differently on each of the sub-processes, and since they
are somewhat suppressed in the SM due to this unitarity cancellation,
they can be enhanced significantly in the presence of anomalous
couplings.

In Fig.~\ref{fig:coupling-scan-nonfact} we show this effect under the
VBF cuts of Eqs.~\eqref{eq:cuts1}--\eqref{eq:cuts2}.\footnote{Here we
apply VBF cuts because the NNLO non-factorizable corrections are
computed in the eikonal approximation which is only strictly valid
when $m^{\textrm{tag}}_{jj} \gg p_{T,j_i}, \, p_{T,H_i}$.} In the SM,
under these cuts, the relative contribution of the NNLO
non-factorizable corrections is $-0.23\%$ (to be compared to the
factorizable NLO$\to$NNLO correction of $-1.7\%$). From
Fig.~\ref{fig:coupling-scan-nonfact} we see that this can grow by more
than $40\%$ to $-0.33\%$. Phenomenologically this has very little
impact, since the corrections are still sub-leading, but it is of some
theoretical interest. Typically, the VBF approximation is justified by
the statement that contributions beyond the factorized approximation
in a strict sense are small.

If non-factorizable
corrections change when anomalous couplings are considered this may
have an impact of the quality of the VBF approximation. We note that
in the recent work of Ref.~\cite{Braun:2025hvr} the authors considered
EW \hhjj{} production retaining all relevant diagrams, including
non-factorizable corrections that appear in the $s$-channel. In figure~3 of that reference they report that the NLO QCD corrections are
\emph{not} entirely flat. We suspect that this is due to the
non-factorizable corrections, and would indicate that the quality of
the VBF approximation deteriorates when anomalous couplings are
considered. We leave a detailed study of this effect for the future.

\section{Conclusions and outlook}
\label{sec:conclusions}
In this article we presented an extension of the \provbfhh tool to
account for non-SM values of the Higgs couplings $\lamhhh$, $\ghhvv$,
and $\ghvv$ in the kappa framework. Moreover, we developed a new
implementation of the VBF-induced $\hhjj$ process in the \PBOX{}. With
this tool fully differential NLO predictions matched to PS programs
such as \PYTHIA{} or \HERWIG{} can be obtained both within the SM and
the kappa framework. We note that this is the first time that NLO+PS
accurate predictions can be obtained with \PYTHIA{}. Having access to
NLO-accurate differential distributions is important to disentangle
the effect of the three anomalous couplings, highlighting the need to
have them implemented in a flexible tool like the one presented here.

Using these tools we systematically explored the impact of
perturbative corrections and parton shower effects for the SM and
found that the NLO+PS results provide a good approximation of the NNLO
predictions for distributions of the tagging jets and Higgs
bosons. Larger differences are found for observables related to
subleading jets.  We then investigated the sensitivity of selected
observables on various Higgs coupling factors and found that non-SM
values compatible with current experimental bounds can result in
distributions differing dramatically from the corresponding SM
expectations.

Finally we also studied the impact of anomalous couplings when
considering non-factorizable corrections to \vbfhh{}. We found that
they are highly sensitive to the exact coupling values, and can be
enhanced by as much as $40\%$ compared to in the SM.

It is worth keeping in mind that our analysis does not take into
account any EW corrections. Since the effect of EW corrections is most
pronounced in tails of distributions, where the effect of anomalous
couplings also tend to show up, it might be important to study their
interplay. We leave that to future studies.

The latest version of the \provbfhh tool can be obtained from
\url{https://github.com/alexanderkarlberg/proVBFH}.  The new NLO+PS
code is made available via the \POWHEGBOXVV{} repository, see
\url{https://powhegbox.mib.infn.it/}.

\section*{Note added}
While finalising this work we were made aware of
Ref.~\cite{Braun:2025hvr}. Our work and that reference are somewhat
complementary, as we have focused on a parton shower matched
implementation of EW Higgs pair production in the VBF approximation,
and compared it to an NNLO computation, whereas Ref.~\cite{Braun:2025hvr}
is at fixed order but goes beyond the structure-function approach. We
leave a detailed comparison of our implementations for future work.

%
\section*{Acknowledgements}
%
We are grateful to Gudrun Heinrich, Jens Braun, Marius H\"ofer, and
Pia Bredt for sharing their draft with us before publication. BJ and
SR acknowledge support by the state of Baden-W\"urttemberg through
bwHPC and the German Research Foundation (DFG) through grant no INST
39/963-1 FUGG. 

%

\bibliographystyle{JHEP}
\bibliography{vbfhh}

\end{document}